\pdfoutput=1

\documentclass[twocolumn, 14pt]{article}
\usepackage[left=2.9cm,right=1.9cm,bottom=4cm,top=2cm]{geometry}

\usepackage{amsmath}
\usepackage{amssymb}
\usepackage{graphicx}
\usepackage{cite}
\usepackage{color} 

\usepackage[labelfont=bf,labelsep=period,justification=raggedright]{caption}

\newcommand{\captionfonts}{\footnotesize}

\makeatletter  
\long\def\@makecaption#1#2{%
  \vskip\abovecaptionskip
  \sbox\@tempboxa{{\captionfonts #1: #2}}%
  \ifdim \wd\@tempboxa >\hsize
    {\captionfonts #1: #2\par}
  \else
    \hbox to\hsize{\hfil\box\@tempboxa\hfil}%
  \fi
  \vskip\belowcaptionskip}
\makeatother   

\makeatletter
\renewcommand{\@biblabel}[1]{\quad#1.}
\makeatother

\date{}

\begin{document}
\thispagestyle{empty}
\twocolumn[
  \begin{@twocolumnfalse}
\begin{flushleft}
\vspace*{-1.5cm}
{\huge
\textbf{Strain-Induced Alignment in Collagen Gels}
}
\\
\vspace*{0.5cm}
{\bf D. Vader$^{1}$, 
A. Kabla$^{2}$,
D. Weitz$^{1,3}$, 
L. Mahadevan$^{1,\ast}$}
\vspace*{0.3cm}
\\
{\footnotesize 
{\bf 1} School of Engineering and Applied Sciences, Harvard University, Cambridge, MA, USA

{\bf 2} Department of Engineering, University of Cambridge, Cambridge, UK

{\bf 3} Department of Physics, Harvard University, Cambridge, MA, USA

}
\end{flushleft}

\section*{Abstract}

Collagen is the most abundant extracellular-network-forming protein in animal biology and is important in both natural and artificial tissues, where it serves as a material of great mechanical versatility. This versatility arises from its almost unique ability to remodel under applied loads into anisotropic and inhomogeneous structures. To explore the origins of this property, we develop a set of analysis tools and a novel experimental setup that probes the mechanical response of fibrous networks in a geometry that mimics a typical deformation profile imposed by cells \textit{in vivo}. We observe strong fiber alignment and densification as a function of applied strain for both uncrosslinked and crosslinked collagenous networks. This alignment is found to be irreversibly imprinted in uncrosslinked collagen networks, suggesting a simple mechanism for tissue organization at the microscale. However, crosslinked networks display similar fiber alignment and the same geometrical properties as uncrosslinked gels, but with full reversibility. Plasticity is therefore not required to align fibers. On the contrary, our data show that this effect is part of the fundamental non-linear properties of fibrous biological networks.
\vspace*{0.3cm}
{\footnotesize 

\textbf{Citation:} Vader D, Kabla A, Weitz D, Mahadevan L (2009) Strain-Induced Alignment in Collagen Gels. PLoS ONE 4(6): e5902. doi:10.1371/journal.pone.0005902
           
\textbf{Received} March 6, 2009; \textbf{Accepted} April 21, 2009; \textbf{Published} June 16, 2009

\textbf{Copyright}: \copyright \  2009 Vader et al. This is an open-access article distributed under the terms of the Creative Commons Attribution License, which permits unrestricted use, distribution, and reproduction in any medium, provided the original author and source are credited.

\textbf{Funding}: This work was supported by NIH Bioengineering Research Partnership grant R01 CA085139-01A2 and by NSF IGERT program in biomechanics. The funders had no role in study design, data collection and analysis, decision to publish, or preparation of the manuscript.

\textbf{Competing Interests:} The authors have declared that no competing interests exist.
\\
$\ast$ E-mail: lm@seas.harvard.edu
}

\vspace*{1cm}

  \end{@twocolumnfalse}
  ]

\section*{Introduction}

Fiber networks, which arise in a range of natural and technological situations, are prime candidates for a wide spectrum of applications requiring tunable mechanical, transport and chemical properties \cite{Star2006}. In nature, these networks self-assemble to generate important structural and functional elements at various length scales: actin, intermediate filaments and microtubules are the main components of the cytoskeleton \cite{Alberts2002}; spectrin confers versatile qualities to red blood cell membranes \cite{Gov2003}; fibrin is an essential element in hemostasis \cite{Weisel2004}; collagen is the main component of the extracellular matrix (ECM) in the animal kingdom \cite{Palsson2003} and cellulose is used by plants to build cell walls \cite{Taiz2002}.

The mechanical function of biological fiber networks is essentially two-fold: (i) at the subcellular (actin, spectrin) and supracellular (collagen, fibrin) scales, the material offers little resistance and high sensitivity to small deformations, allowing it to be easily remodeled locally ; (ii) at larger strains it stiffens strongly to ensure cell and tissue integrity \cite{Sacks2000}. The non-linear stiffening, while  observed in many biological systems \cite{Storm2005, Janmey2007}, is not fully understood yet, with theories focusing on one of two broad mechanisms: (i) microstructural nonlinearities of individual filaments \cite{Storm2005}, and (ii) collective non-affine deformations of multiple filaments \cite{Heussinger2006a, Wyart2008}. To unravel the relative importance of these mechanisms, a range of experimental tools have been developed  to quantify the network's mechanical non-linearity in systematic ways and relate the material micro-structure (network density and morphology, fiber behavior) to the mesoscopic stress-strain laws. These tools fall into two broad categories: simple shear in cone-plate or parallel plate geometries, and uniaxial/biaxial stretch.

Simple shear deformations are commonly used to study purified protein networks. This technique requires low sample volumes and provides a consistent set of experimental tools and generic protocols to probe the visco-elastic properties of soft gels in both the small-strain (linear) and large-strain (non-linear) regimes, and in addition, normal stresses can be measured. Recent data collected by {\it Janmey et al.} \cite{Janmey2007} show in particular that sheared biopolymers exert negative normal forces, a fact that is in contradiction with the hyperelastic behavior of other well studied elastomers. The broad availability of experimental data in that geometry has encouraged a large number of related theoretical and numerical studies \cite{Head2003, Wilhelm2003, Onck2005}, focused primarily on the linear response of the material. However, since simple shear rheology assumes that the material undergoes purely isochoric deformations in the limit of small strains, it only allows for partial exploration of material behavior. In particular, these experiments do not allow one to study completely the non-linear regime (strain typically larger than 10\%) that is most relevant in many biological situations (single cell or tissue deformation). And furthermore, it does not allow for a probe of the dilatational rheology of the networks.

In contrast, at mesoscopic scales, uniaxial and biaxial testing are most common for tissue mechanical characterization \cite{Billiar1997, Sacks2000, Ivancic2007, Cheng2007} and have been used to study reconstituted collagen networks \cite{Knapp1997, Tower2002, Girton2002, Roeder2004, Krishnan2004}, the simplest  tissue equivalents \cite{Tranquillo1999}. In contrast with simple shear, uniaxial stretch generically leads to non-isochoric deformations, and hence allows one to measure quantities such as the material Poisson ratio  which can have values as large as $\nu \approx 3$ for strongly deformed collagen gels. These values arise in highly anisotropic materials, as reported for instance for solid foams \cite{Lee1997}, and it is somewhat surprising to see similar behavior in  \textit{in vitro} collagen gels which display little or no anisotropy in their undeformed state.

\begin{figure}[!htb]
  \centering 
  \includegraphics[width=7.5cm]{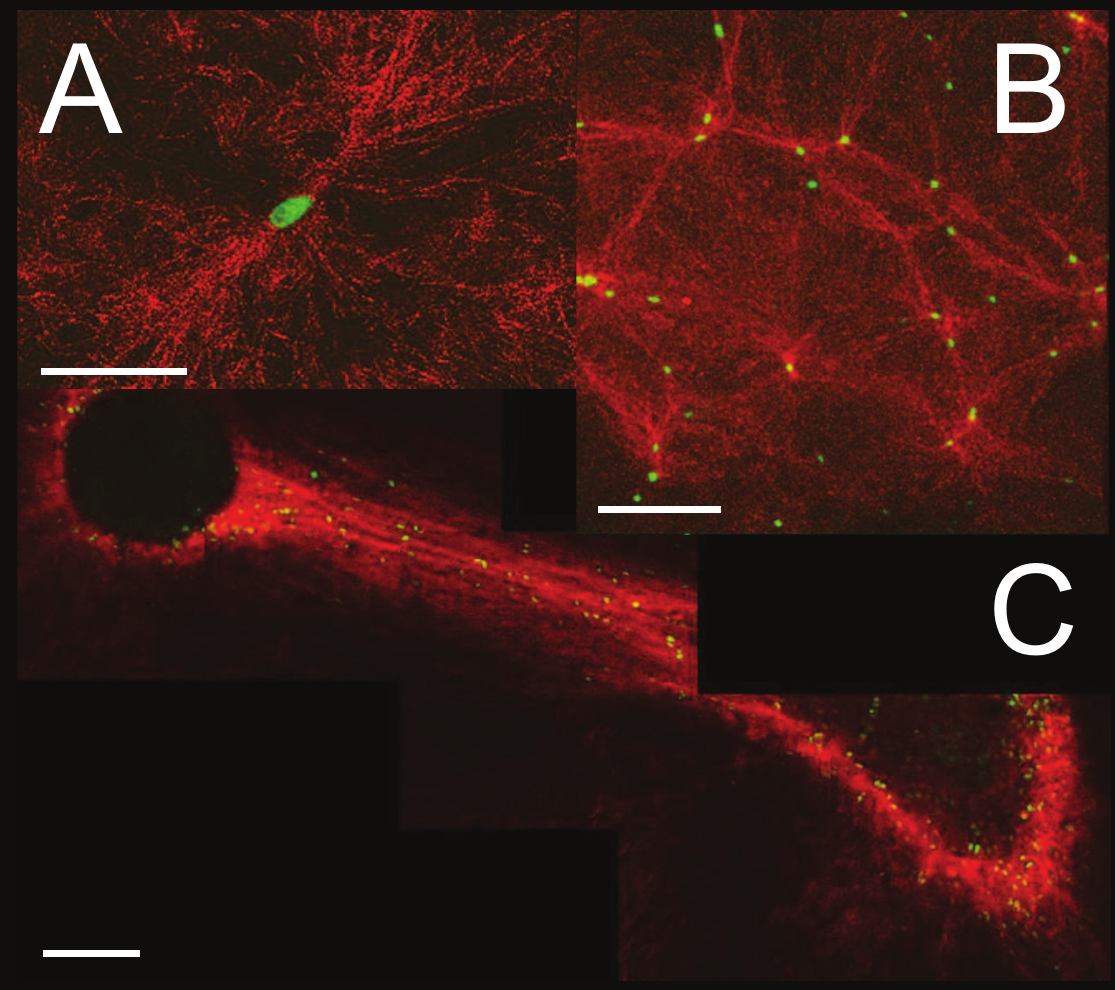}
  \caption{ {\bf Collagen gel morphological changes induced by presence of cells.} (A) Single U87 glioblastoma cell in a collagen network 10 hours after gel polymerization. bar=50 $\mu$m. (B) Several U87 cells on the surface of a collagen gel 10 hours after gel polymerization. bar=200 $\mu$m. (C) Two cell colonies embedded in a collagen matrix 48 hours after gel polymerization. bar=200 $\mu$m. Fibers (artificial red color) are imaged through confocal reflectance; cell nuclei (green) are labeled with a GFP-histone heterodimer.}
  \label{fig:alive}
\end{figure}

To understand this, we recall that early studies \cite{Harris1980} on cell/matrix interactions show that cells or groups of cells tend to generate tensile forces on the extracellular environment . When cell colonies were plated on fibrous materials such as collagenous gels, Harris and Stopak reported the formation of anisotropic and denser regions connecting these cellular assemblies, and showed that the matrix structure has a strong influence on cell motility. Although these observations are well accepted, little is known about the mechanical response of a fibrous matrix subject to an internal local strain. Neither of the mechanical characterizations described previously focus on how deformation changes the microstructure at the fiber scale, an issue of particular importance in the large strain regime, that is all too easy to observe (Figure \ref{fig:alive}).

In this paper, we use collagen type I gels as a model system to address this question and shed light on the morphological evolution of both the fiber \footnote{Although the expression "collagen fiber" traditionally refers to large bundles of collagen fibrils, we will use the words "fiber" and "fibril" interchangeably in this paper to refer to the 0.5 micron diameter bundles.} and the network on an externally imposed stretching strain. Collagen is a convenient biomaterial for biomechanical studies for a number of reasons: a) it is readily available in large amounts, which makes it suitable for milliliter-size gels; b) {\it in vitro} reconstituted networks have fibers that are easily identifiable using confocal microscopy; c) many of its properties have been extensively studied \cite{Raub2007, Gentleman2003, Roeder2002, Brightman2000, Bozec2005, Knapp1997}; d) the large diameter of the fibers ($\approx 0.5 \mu m$ for collagen fibrils \cite{Roeder2004}) and the stability of the network \cite{Leikina2002, Persikov2002} make it easy to handle and image over a range of spatial and temporal scales; e) fibrillogenesis is conveniently controlled \textit{in vitro} by pH, temperature and concentration \cite{Roeder2002, Raub2007}.

We first verified the presence of cell-induced alignments and densification with our experimental system. As shown in figure \ref{fig:alive}A, an isolated human glioblastoma cell (see Methods) in a collagen network induces stress variations and modifies the network texture in its vicinity \cite{Friedl1998, Kim2006}. Several isolated cells on the surface of a collagen gel produce fiber alignment and network densification along lines connecting individual cells (figure \ref{fig:alive}B). Following Stopak and Harris, we also observed fiber alignment on macroscopic length scales when  we introduce large cell assemblies in the same extra-cellular environment (figure \ref{fig:alive}C). Since active matrix remodeling is restricted to the vicinity of living cells, such an effect can only be accounted for by the mechanical properties of the network. 

With these observations in mind, we  employ a specific experimental approach and develop a set of tools to quantitatively study the coupling between strain and the morphology of fibrous networks in a range of strain and strain rates that are typical of many biomechanical situations. Experiments on cell colonies suggest that such a process can be conveniently studied at the millimeter scale, and over a time-scale of a few hours. However, instead of using cells to deform the extra-cellular matrix, we use an external imposed displacement to stretch collagenous samples and monitor the gel response. In particular, this experiment allows us to probe a range of dynamical regimes and independently tune the biochemistry (crosslinking) to study the coupling of tensile strain to network density and fiber orientation in a controlled setting and investigate the origin and generality of these mechanical processes. This also allows us to address the outstanding question of the mechanical reversibility of these patterns in an extracellular environment.

\section*{Methods}

\subsubsection*{Network synthesis}

\textit{In vitro} collagen networks are prepared according to a previously described cell culture-compatible protocol \cite{Kaufman2005}, with a final collagen concentration ranging from 0.5 to 4.0mg/mL. Solutions consist of 10\% 10X minimum essential medium (Invitrogen, Carlsbad, CA), 10\% fetal bovine serum (JRH
Biosciences, Lenexa, KS), 1\% penicillin-streptomycin (Invitrogen), bovine collagen diluted to desired concentration (from 3.1mg/mL or 6.4mg/mL batch, Inamed Biomaterials, Fremont, CA), a few $\mu$L of 1M sodium hydroxide (NaOH, Sigma, St. Louis, MO) to bring pH to neutral, 50mM sodium bicarbonate (NaHCO$_3$, Sigma) buffer and deionized water. 800$\mu$L of solution are pipetted onto glass-bottom Petri dishes (MatTek, Ashland, MA) Samples then polymerize for 30-60 minutes in a cell incubator at 37$^\circ$C, 5\% CO$_2$. After polymerization, samples are 20mm in diameter and 1-2mm in height.

In addition to untreated \textit{in vitro} collagen gels, we also prepare polymerized samples, to which we add glutaraldehyde (GA) - a common cell and tissue fixative. The effect of this is an increase in the overall stiffness of the gel by at least an order of magnitude (from a few tens of Pascals to over 1kPa for a 1mg/mL gel, as evaluted using a cone-plate rheometer), without noticeably changing the structure of the collagen network (fiber width and length, mesh size). 2mL of 4\% v/v GA (Sigma, St. Louis, MO) in water is pipetted onto the sample, which is then incubated once more for at least 2 hours. It is subsequently rinsed twice with deionized water. Before use, all samples are immersed into 2mL of deionized water, to allow the gel to swell, and to reduce friction.

\subsubsection*{Cell experiments}
U87 human glioblastoma cells are cultured as described in \cite{Kaufman2005}. After passaging the cells, tissue equivalents are generated by diluting the cells to approximately $10^5$/mL in an unpolymerized collagen solution at 1.5mg/mL. After polymerization in a 37$^\circ$C temperature- and humidity-controlled incubator, the spacing between individual cells, as seen in figures \ref{fig:alive}a and \ref{fig:alive}b, is on the order of 100$\mu$m. Cell colonies (or spheroids), with an estimated $10^3$ cells, are generated with the hanging droplet method \cite{Kelm2003} and subsequently seeded in 500$\mu$L of collagen solution at 1.5mg/mL shortly before it polymerizes.

\subsubsection*{Bulk rheology}

We use an AR-G2 (TA Instruments, New Castle, DE) rheometer with a 4$^\circ$, 40mm cone-plate geometry with a 109$\mu$m gap. 1.2mL of collagen solution is pipetted onto the 37$^\circ$C preheated bottom plate of the rheometer and the cone is lowered onto the sample. We use a solvent trap to prevent the sample from drying during the measurement. During polymerization, the increase in G$'$ and G$''$ is probed by continuously oscillating the sample at a fixed 0.5\% strain amplitude and at a frequency of 0.2Hz. The oscillatory strain sweep is performed at the same frequency and temperature, after the gel has polymerized for 2-3hrs. The strain amplitude is increased logarithmically  until the sample breaks.

During oscillatory strain sweeps, we simultaneously record the maximum stress and strain of the sample for each oscillatory cycle. To characterize the onset of stiffening from the stress-strain data, we define the critical strain as the value at which the stress $\sigma$ exceeds the product $G'_0\gamma$ by more than 10\%, where $G'_0$ is the elastic modulus in the small-strain linear regime.

\subsubsection*{Mechanical setup}

\begin{figure}[!htb]
 \centering
 \includegraphics[width=8cm]{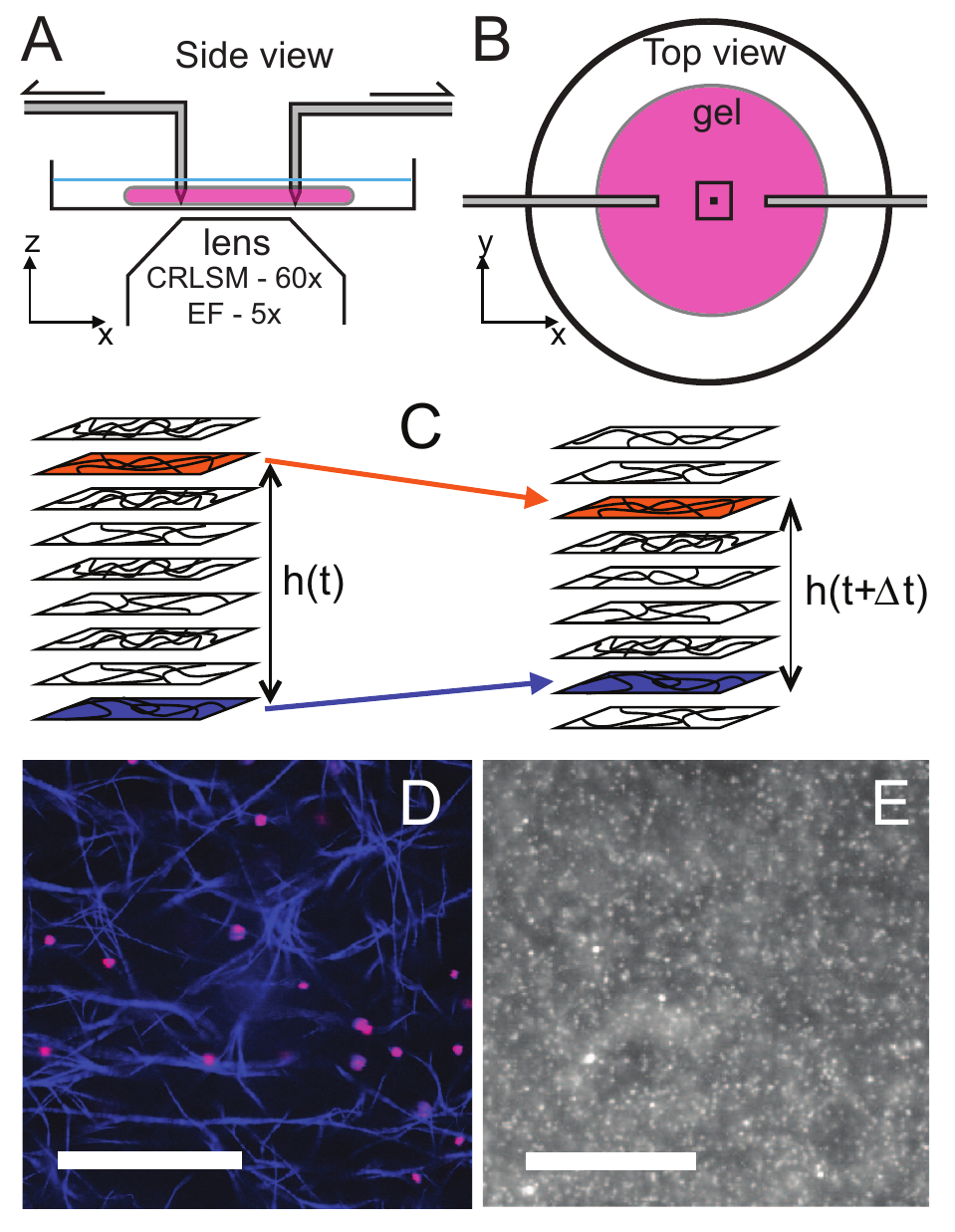}
 \caption{ {\bf Mechanical setup and sample imaging.} (A) Side and (B) top views of the mechanical setup used to deform the network; the collagen gel has a pancake-like shape, typically 1mm in thickness and 20mm in diameter. As defined in our experiments, the stretch axis is $x$. Drawn to scale, the two squares represent the fields-of-view of the wide-field fluorescence images (5x) and confocal reflectance images (60x). (C) Correlation of multiple slices over time gives an estimate of the interslice distances, and hence vertical strain. (D) Collagen network (blue) obtained with confocal reflectance. Fluorescent tracers (pink) are embedded in the network. Scale bar 20$\mu m$. (E) Wide-field fluorescence image of the embedded beads. Scale bar 500$\mu m$.}
 \label{fig:methods}
\end{figure}

We place a polymerized collagen sample onto a glass cover slip and perforate it with two rough-ended 1mm-diameter glass cylinders (capillaries) (see figures \ref{fig:methods}A,B), which are gently pushed all the way to the glass bottom to prevent the collagen from slipping beneath them. Each cylinder is attached to two secondary transverse elastic capillary rods, themselves attached to two linear transducers (Newport, Irvine, CA) controlled by the ESP300 controller (Newport). The transverse capillaries act as springs that allow to maintain contact with the bottom cover slip of the dish with constant pressure. The tips, initially 1cm from each other, can then be moved apart at speeds ranging from 0.125 to 12.5 $\mu m/s$, corresponding to strain rates of 2.5$\cdot 10^{-5}$ to 2.5$\cdot 10^{-3}$ per second; this range includes measured rates of cell-induced contraction \cite{Pizzo2005}. The movement of the tips results in the local stretching of the gel sitting between them. For imaging purposes, the whole mechanical setup (motors and tips) is clamped to the microscope sample holder plate.

\subsubsection*{Confocal imaging}

We use a Zeiss LSM 510 Meta (Carl Zeiss Microimaging Inc., Thornwood, NY) equipped with a 488nm Argon laser line and several photomutliplier tubes. We set the Meta channel of the microscope (which allows for selection of specific wavelengths) to detect wavelengths between 474 and 494nm to allow for the fact that we work in reflectance mode \cite{Gunzer1997}, which has the significant advantage of avoiding the use of fluorescent dyes in the samples. A 60X 1.2-NA Olympus water immersion objective (Olympus America Inc., Center Valley, PA) is mounted onto the microscope.

While deforming the sample, we acquire timelapse 2D confocal images at various heights between 50 and 150$\mu$m from the bottom surface. Tracking in-plane deformations in multiple slices improves the statistics of our analysis; moreover, tracking out-of-plane motion via image correlation allows us to estimate the vertical deformation of the sample (figure \ref{fig:methods}C). The timelapse interval is 10s, which corresponds to 0.25\% imposed deformation at the typical strain rate.

\subsubsection*{Image segmentation and fiber detection}

Fiber orientation is calculated for each 2D confocal slice by several image processing steps: the raw images are filtered using a 2D Gaussian blur and subsequently thresholded so that at 0\% stretch, 10\% of the pixels are above that threshold; this threshold value is applied to all subsequent images of the same experiment. A circular window of diameter 15px moves across the thresholded image, and the 2nd order moment tensor $M$, defined below, is locally calculated using binary pixel weights: below the threshold level, the pixel weight $A_{ij}$ is 0, above that level, $A_{ij}$ is 1. Typically, the moment tensor quantifies the spatial distribution of weight around a center of mass and its eigenvalues give an indication as to whether weights are distributed isotropically around the center or not. Similarly here, $M$ quantifies the distribution of pixel intensities around the intensity-weighted center of mass, with $X_{ij}$ and $Y_{ij}$ the pixel coordinates in the local circular window:

\begin{equation}
 M \! = \! \! \! \\
\left[ \! \! \! \! \begin{array}{cc} 
\ \! \! \! \scriptstyle \sum A_{ij} (X_{ij} - \bar{X})^2 &
\ \! \! \! \! \! \! \! \! \! \! \scriptstyle \sum A_{ij} (X_{ij} - \bar{X}) (Y_{ij} - \bar{Y}) \\
\ \! \scriptstyle \sum A_{ij} (X_{ij} - \bar{X}) (Y_{ij} - \bar{Y}) &
\ \! \! \! \! \! \! \! \! \! \! \scriptstyle \sum A_{ij} (Y_{ij} - \bar{Y})^2
\end{array} \! \! \! \! \right]
\end{equation}

\begin{equation}
r = \frac{\hbox{Max} \left\{  \hbox{Eigenvalues} (M) \right\}}
 {\hbox{Min} \left\{  \hbox{Eigenvalues} (M) \right\}}
\end{equation}

The ratio $r$ characterizes the aspect ratio of binarized image fragment enclosed in the sliding window and can be used to detect fibers: a single fiber passing through the middle of the window, with a diameter less than half the window size will yield a high aspect ratio; a single fiber in the corner of the window or multiple fibers in the window will give a low aspect ratio. We only keep the regions with $r > 2$ for which the eigenvector corresponding to the higher eigenvalue of $M$ provides the local fiber orientation $\phi$.

The choice of window size (15 pixels) is a compromise between increasing angular resolution at low strains and avoiding multiple fibers in the window. As shown later on, the fiber density strongly increases at high strain, resulting in less accuracy in the orientation analysis. In practice, this sets the limits of the method. See {\it Methods S1, Figure S1 and Figure S2} for validation of this image processing algorithm.

\subsubsection*{Orientational order parameter}

To quantify the network anisotropy as a function of the applied deformation, we calculate an orientation tensor $\Omega$ and an associated nematic order parameter $\mu$ from the distribution of the fiber orientation $\phi$ at each time step, defined by:

\begin{equation}
\Omega = \left( \begin{array}{cc}
\left< \cos ^2 (\phi) \right> & \left<\cos\phi \sin\phi  \right>\\
\left< \cos\phi  \sin\phi  \right> 				& \left<\sin ^2 \phi \right>  \end{array}
 \right)
\end{equation}
\begin{equation}
\mu = \hbox{Max} \left\{  \hbox{Eigenvalues} (2 \;\! \Omega - \hbox{\bf Id})  \right\} 
\end{equation}
where $\bf Id$ is the identity matrix, and $< \cdot>$ denotes spatial averaging over the domain of interest. The order parameter ranges from $\mu =0 $ for a uniform angular (isotropic) distribution to $\mu =1 $  for a perfectly aligned system. Although this parameter, based on a 2D image analysis, only characterizes the order in the horizontal plane, it does provide a suitable signature of the microstructure evolution, and in particular its non-linear behavior.

\subsubsection*{Deformation field at the mesoscale}

We also characterize, at the mesoscopic millimeter scale, the strain field $\varepsilon_{xx}$, $\varepsilon_{yy}$ and $\varepsilon_{zz}$ induced by the imposed displacement of the glass tips. Vertical strain is estimated by following the individual displacement of confocal z-stacks. For each slice of a stack taken at a time $t$, this is done by calculating  the correlation with the slices obtained at a neighboring time $t+\Delta t$ (see figure \ref{fig:methods}C). The height of the slice which provides the largest correlation value indicates the new location of the material layer and this information is used to calculate the vertical component of the strain.

The deformation in the $xy$ plane is measured by a PIV (particle imaging velocimetry) method. 1$\mu m$ diameter rhodamine Fluospheres (Invitrogen, Carlsbad, CA), at a volume ratio of 1:100000, are used as tracers to measure the deformation of the gel at millimeter length-scales. Most of the carboxy-coated particles stick to the network, as seen in figure \ref{fig:methods}D. Using a 5x lens on a Zeiss wide-field microscope and focusing on the middle of the sample, we image the local density of these particles, which displays heterogeneities as seen in figure \ref{fig:methods}E. An image cross-correlation technique is then used to track these heterogeneities at a scale of $10-50 \mu m$ as the network is progressively deformed. To identify the local displacement of a mesoscopic region of the gel located at $(x,y)$ from time $t$ to $t+\Delta t$, we extract a domain of 48x48 pixels surrounding $(x,y)$ at $t$ and look for the best matching region - i.e. the one that maximizes the cross-correlation function - in the image obtained at $t+\Delta t$. The cross-correlation function is defined as:

\begin{equation}
\rho_{AB} = \sum{(A_{ij} - A_{avg})(B_{ij} - B_{avg}) / (\sigma_A \sigma_B)}
\end{equation}
where A and B are the pixel intensity values associated with the two regions of interest, $A_{avg}$ and $B_{avg}$ are the local average pixel intensities, and $\sigma_A$ and $\sigma_B$ are the standard deviations of intensity values of those regions.

This tracking through cross-correlation process is iterated over time to extract the full trajectory of the material point and corresponds to a Lagrangian description of the material. For a grid deformation that is fairly homogeneous in the field of view of the microscope (millimeter scale), we use a least-squares planar fit of the nodal displacement to quantify the material deformation at the mesoscale. The deviation from the fit provides a measure of the error on the deformation field.

{These measurements allow us to extract a number of strain and stress characteristics. In particular, in our geometry, the normal stresses $\sigma_{yy}$ and $\sigma_{zz}$ are negligible as the gel is not attached on the lateral sides, and this allows us to estimate the incremental Poisson ratio, defined as:
}    
\begin{equation}
\nu_{xy} = - \partial \varepsilon_{yy} / \partial \varepsilon_{xx}
\label{Poisson_define}
\end{equation}
{\noindent which characterizes the coupling of incremental deformations in orthogonal directions. In two dimensions and in the small deformation limit, the area of the grid $A(\epsilon_{xx})$ is related to the Poisson ratio via the following relationship:
}

\begin{equation}
\frac{1}{A(0)} \frac{dA}{d\epsilon_{xx}} = 1 - \nu_{xy}
\end{equation}
{Most materials respond to tension with a slight increase in their area (volume in three dimensions), which for an isotropic material translates into the condition:
}

\begin{equation}
\nu_{xy} \le 1
\end{equation}

The analogous condition in 3 dimensions is $ \nu_{xy} \le 0.5$. At large deformations for an isotropic material, the criterion is slightly more complex, but the critical strain, beyond which the change of area (or volume in 3D) with respect to elongational strain becomes negative, remains of the same order in practice.

\section*{Results}

\subsubsection*{Rheological characterization of the samples}

\begin{figure}[!htb]
  \centering
  \includegraphics[width=7.5cm]{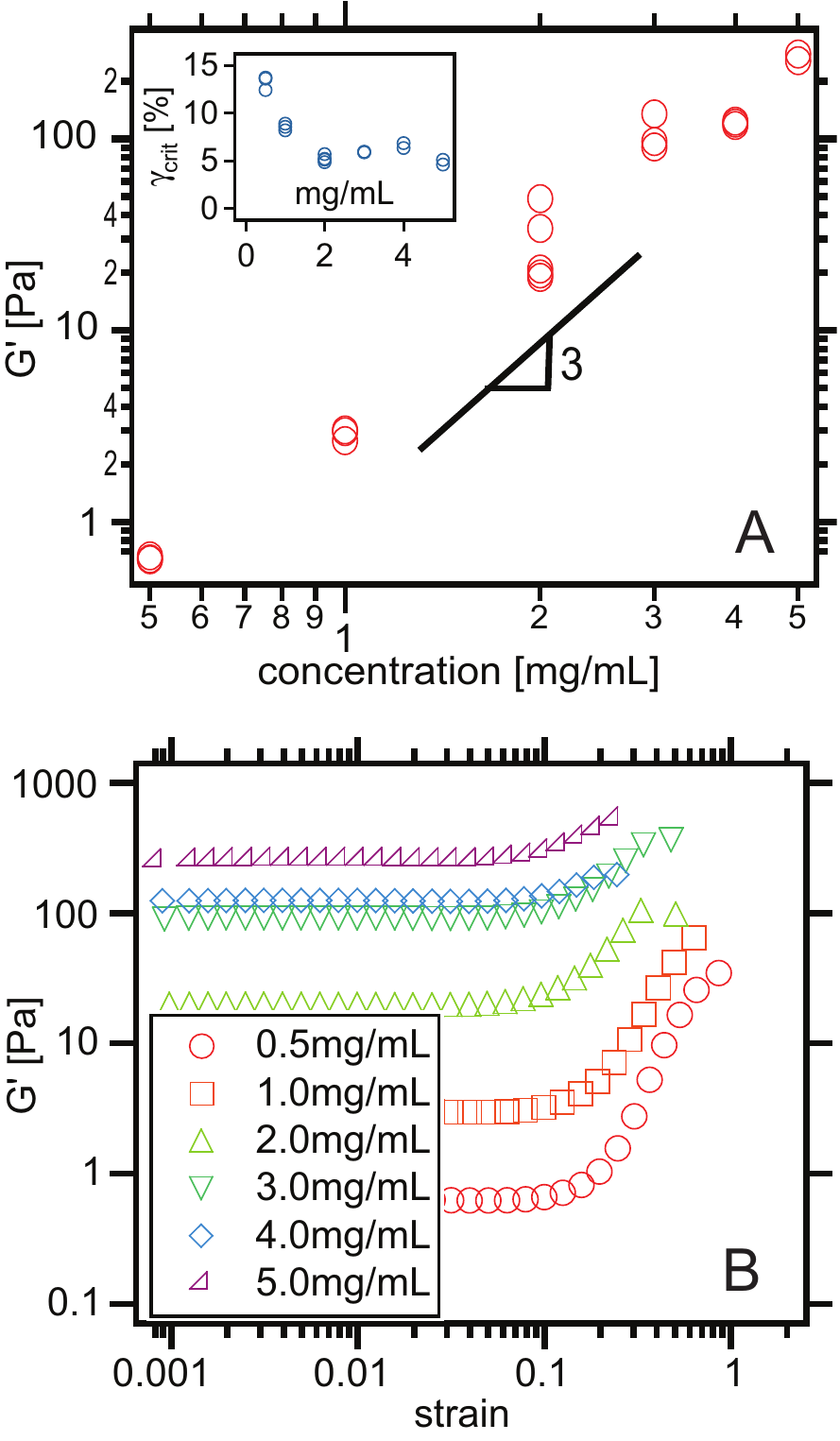}
  \caption{ {\bf Bulk rheology measurements.} (A) Linear elastic modulus as a function of collagen concentration. A power-law of 3 is shown for comparison. Inset: the onset of non-linear mechanical behavior, as defined in Methods. (B) Elastic modulus $G'$ during oscillatory strain sweeps and as a function of collagen concentration.}
  \label{fig:rheology}

\end{figure}

The range of collagen concentrations we work with (0.5-4.0 mg/mL) display mesh sizes from 1-5 $\mu$m (measured through analysis of confocal reflectance slices as in \cite{Kaufman2005}) and span over two orders of magnitude in shear modulus (see figure \ref{fig:rheology}A), with values in close agreement with previously reported data \cite{Kaufman2005, Parsons2002, Velegol2001}.  The linear shear modulus $G'$ (measured at small deformation) has a strong dependence on the concentration $c$, with a behavior well approximated by $G' \sim c^3$ over the probed range. The onset of nonlinear strain-stiffening typically occurs at strains of the order of 5\% (see figure \ref{fig:rheology}B and \cite{Roeder2002}) and has only a weak dependence on collagen concentration (see figure \ref{fig:rheology}A inset).

\begin{figure*}[!htb]
  \centering  \hspace*{-0.15cm}
  \includegraphics[width=17cm]{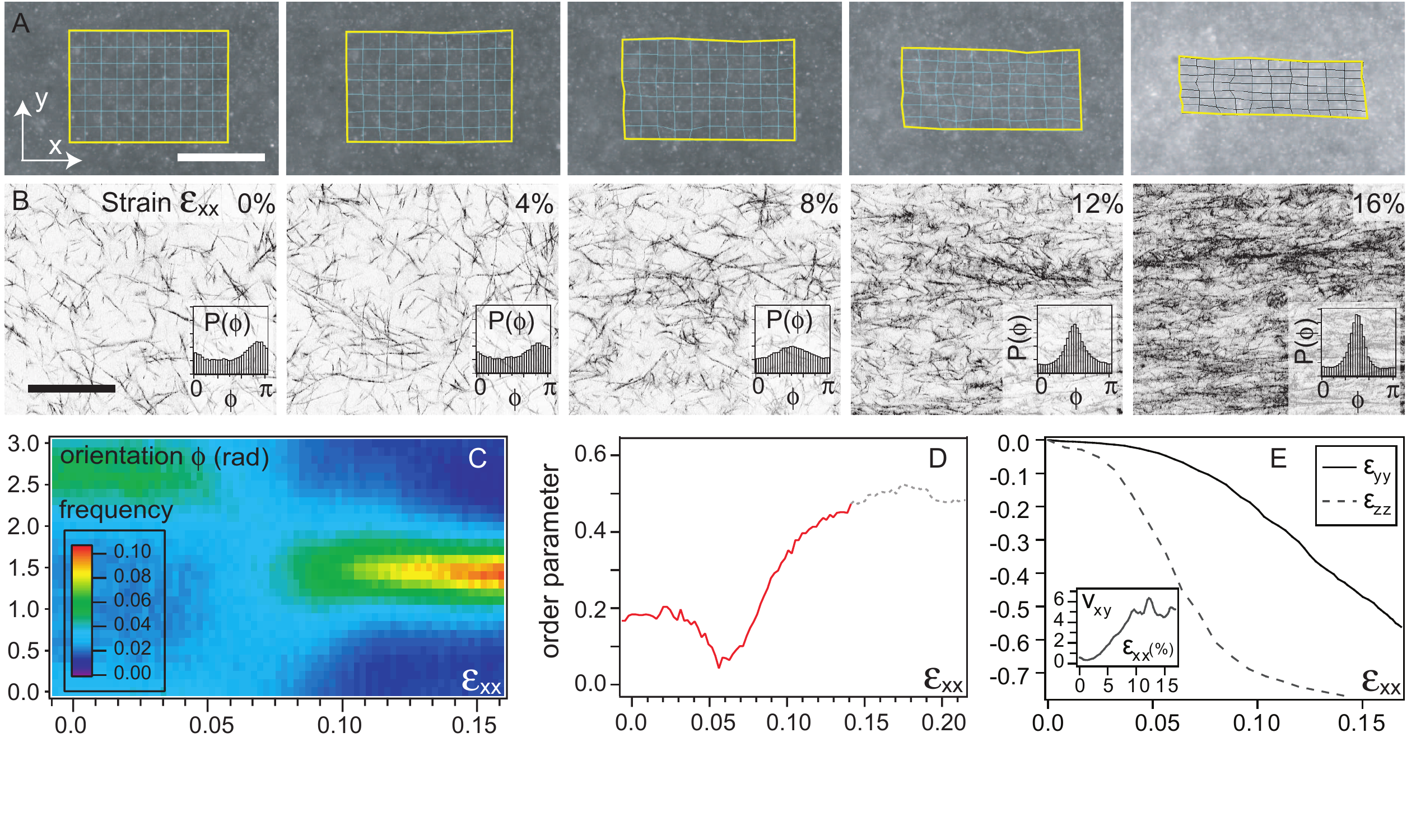}
  \caption{ {\bf Typical results of a sample stretching experiment at micro- and meso-scale.}
The two montages of 5 images each show, for two different 1mg/mL samples and at various strains, (A) wide-field fluorescence images of beads embedded within the network - the super-imposed grid is the result of the tracking of the deformation field (scale bar 500$\mu$m); (B) direct imaging of the fibers through CRLSM (scale bar 50$\mu$m). In inset, each corresponding $\phi$ histogram, with angle values going from 0 to $\pi$. For the same samples depicted above: (C) represents the evolution of the orientation statistics; the color at each point corresponds to the relative count of fibers oriented along a specific direction at a given strain $\varepsilon_{xx}$. (D) shows the order parameter $\mu$ resulting from the data in (C); the curve beyond 15\% stretch is grayed out due to the lack of confidence of the order parameter when the high value of the density prevents a proper detection of the fibers (see Methods). (E) gives the deformations $\varepsilon_{yy}$ and $\varepsilon_{zz}$ as a function of the local strain $\varepsilon_{xx}$. In inset, the incremental Poisson ratio $\nu_{xy}$ as a function of the imposed deformation.}
  \label{fig:fresque}
\end{figure*}

\subsubsection*{Fiber alignment and non-linear Poisson effect}

Figures \ref{fig:fresque}A and \ref{fig:fresque}B illustrate the evolution of the network microstucture over the domain as it is stretched. At low strains ($< 5\%$), no particular alignment can be observed; however, above this threshold both fiber alignment and network density increase. This is quantified in the figures \ref{fig:fresque}C and \ref{fig:fresque}D where we show, as a function of the applied strain, the probability distribution of local in-plane fiber orientation and the resulting order parameter. 

\begin{figure}[!htb]
  \centering
  \includegraphics[width=8cm]{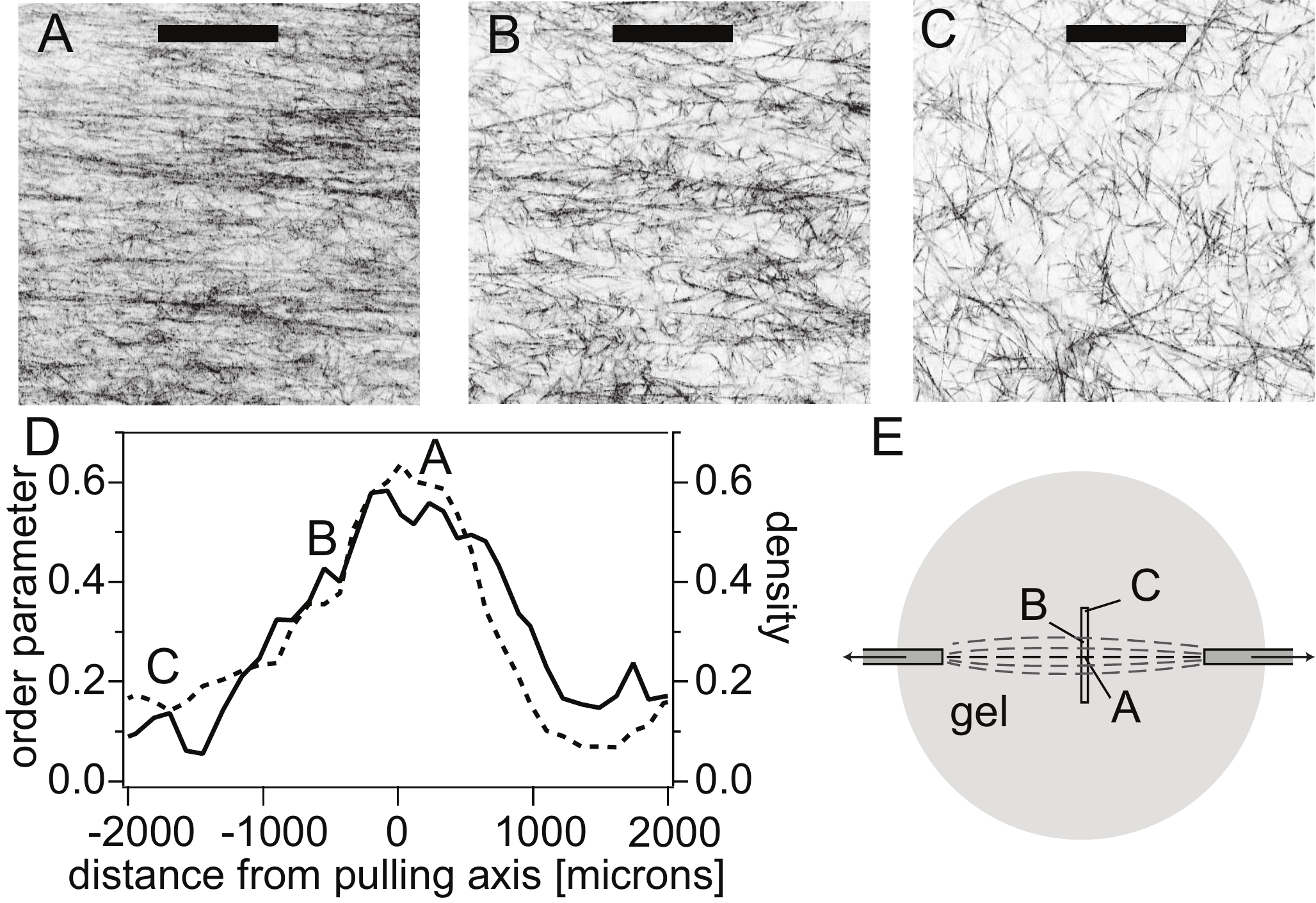}
  \caption{ {\bf Characterization of the spatial variations of collagen fiber alignment and densification.}
(A-C) Confocal reflectance images of 1mg/mL collagen network stretched up to 15\%; images are located at (A) 0, (B) 0.5 and (C) 2 millimeters from the stretching axis. bars = 50$\mu m$. (D) Order parameter and density as a function of distance from the stretching axis, for the same sample. Letters A, B and C on figures (D) and (E) refer to corresponding images above. (E) Drawn to scale, locations of images (A), (B) and (C) with respect to stretching axis.}
  \label{fig:pano}
\end{figure}

For a prescribed displacement of the capillary tips, we also characterized the gel microstructure as we move away from the axis connecting the two capillaries. Both the alignment and fiber density (expressed as fraction of pixels above a given threshold) decrease (figure \ref{fig:pano}). This picture is, as expected, in direct agreement with the alignment pattern induced by cell colonies pulling on extra-cellular matrix shown in figure \ref{fig:alive}, where fiber alignment and density are maximum along the axis joining the colonies, and decay away from it.

In figure \ref{fig:fresque}E we show the average  induced strain components  $\varepsilon_{yy}$ and $\varepsilon_{zz}$ as a function of the externally applied strain  $\varepsilon_{xx}$. When $\varepsilon_{xx} \le  2\%$ there is little contraction in the transverse direction so that the Poisson ratio $\nu_{xy} \sim 0$. As the applied strain increases, the material first thins by contracting in the $z$ direction when  $2\% \le \varepsilon_{xx} \le 5\%$, and only when $\varepsilon_{xx} \ge 5\%$ does it also contract in the transverse $y$ direction, with observed values of $\nu_{xy}$ as high as 5 (figure \ref{fig:fresque}E inset). The lag in response between these two directions can be attributed to the sample geometry as well as a slight initial anisotropy of fiber orientations in the $yz$ plane \cite{Roeder2004}; here we consider only the properties of the planar projection of the network. The large in-plane incremental Poisson ratio $\nu_{xy}$  quantifies the change in local fiber density and is consistent with the confocal observations of densification. In order to quantify and compare these changes with an independent measure of the geometry of deformation, we define a critical strain $\epsilon_{crit}$ for which $dA(\epsilon_{crit})/d\varepsilon_{xx}=0$, beyond which the areal strain (yellow frame on figure \ref{fig:fresque}A) starts to decrease with the applied strain. We find that  $\epsilon_{crit} \sim 5\%$, consistent with the critical strain observed for fiber alignment.

\begin{figure}[!htb]
  \centering
  \includegraphics[width=8cm]{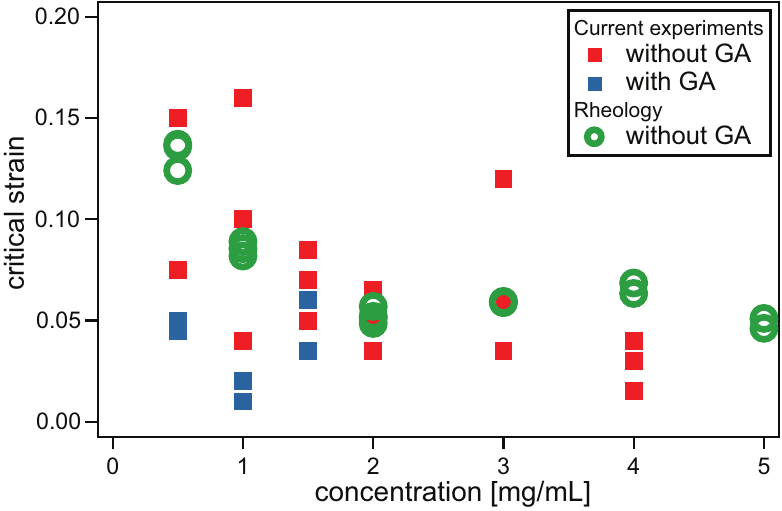}
  \caption{ {\bf Comparison of critical strains measured from bulk rheology and 2-point stretching.}
Comparison of critical strain measured from the PIV method ($\epsilon_{crit}$ such that $dA(\epsilon_{crit})/d\varepsilon_{xx}=0$, see Results) on locally stretched samples with the onset of strain stiffening ($\gamma_{crit}$ such that $\sigma > 1.1 G'_0\gamma$, see Results) obtained from rheological measurements.}
  \label{fig:crossover}
\end{figure}

The critical deformation, as defined above from the kinematic behavior in local stretching tests, can be directly compared with the strain associated with the mechanical stiffening measured in rheological experiments (see figure \ref{fig:rheology}). These two quantities, measured independently, show good correlations in their values and trends. For unfixed collagen samples, critical strain values, measured either from rheological meansurements or from the kinematics, range from a few percents at high collagen concentration to 15 \% at low concentration (see figure \ref{fig:crossover}). The critical strain is therefore very weakly dependant on the concentration, in particular compared with the variation of the elastic modulus at small deformation that varies over more than two orders of magnitude in the same concentration range. GA-fixed samples, whose stiffness is estimated at an order of magnitude higher than their non-fixed counterparts, also show similar values for $\epsilon_{crit}$. Taken together, these results show that the strains above which the gel behavior becomes non-linear as evidenced i) from the elastic modulus for a simple shear geometry and ii) from the Poisson ratio in the local stretching experiments are related with each other and only weakly sensitive to physical scales such as the actual value of the elastic modulus.

\subsubsection*{Orientational ordering is an elastic effect}

\begin{figure*}[!htb]
  \centering  \hspace*{-0.3cm}
  \includegraphics[width=17cm]{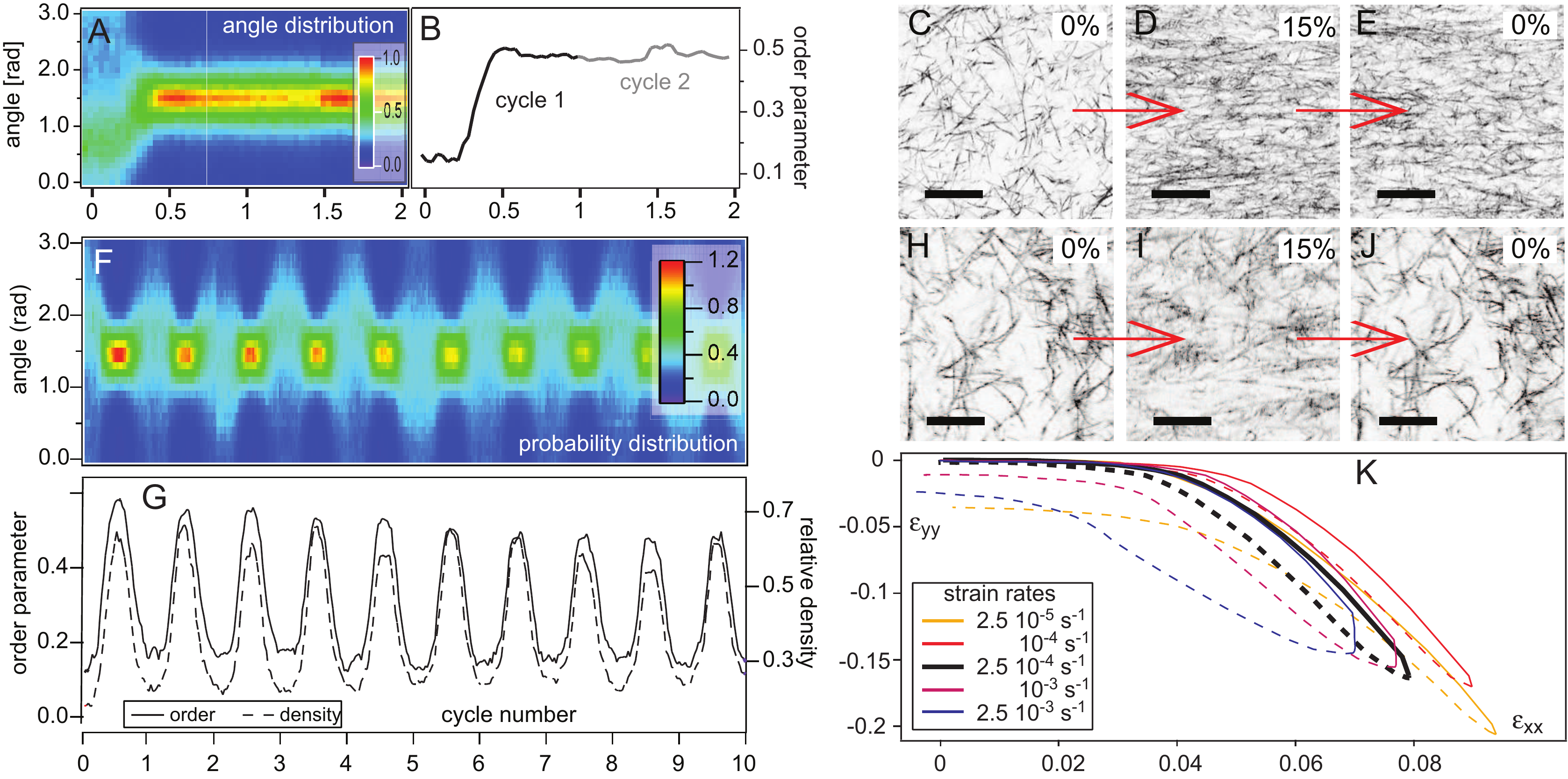}
  \caption{ {\bf Characterization of the reversibility of alignment and densification in crosslinked and uncrosslinked collagen samples.}
Response to cycles of deformation for untreated collagen gels (A-E) and gels treated with glutaraldehyde (F-K). (A) Histograms of the fiber orientation as a function of the imposed strain for untreated sample; color at each point corresponds to the relative count of fibers oriented along a specific direction at a given strain $\varepsilon_{xx}$. (B) Resulting order parameter for untreated sample. (C-E) Confocal reflectance images of a 1mg/mL collagen sample cycled 4 times up to 15\% stretch and back to 0\%, showing the extent of the reversibility at the microstructural level: (C) beginning of 1st stretch cycle; (D) middle of 1st  stretch cycle; (E) end of 1st stretch cycle; bars = 50$\mu m$. (F) Histograms of the fiber orientation as a function of imposed strain for glutaraldehyde-treated sample; color-coding as in (A). (G) Resulting order parameter and density for glutaraldehyde-treated sample. (H-J) Confocal images of a 1mg/mL collagen samples with glutaraldehyde, cycled 4 times up to 15\% stretch and back to 0\%: (H) beginning of 3rd stretch cycle; (I) middle of 3rd stretch cycle; (J) end of 3rd stretch cycle; bars = 25$\mu m$. (K) Mesoscopic response to 15\% strain cycles for various strain rates at a constant imposed strain amplitude. The bold line corresponds to the strain rate used for all other experiments.}
  \label{fig:reversibility}
\end{figure*}

Strain-induced alignment arises {\it a priori} from a combination of reversible elastic effects and irreversible inelastic effects.  To disentangle these two contributions, we apply repeated strain cycles to the sample. All pure type-I collagen samples display very little reversibility regardless of their concentration once the imposed  strains exceed about 10\%; the gel never recovers its initial configuration once the capillary tips return to their initial location. In figures \ref{fig:reversibility}A,B we show the evolution of the fiber orientation histograms as the material is cyclically stretched up to strains of 15\%: fiber aligment is permanently imprinted (\ref{fig:reversibility}C-E).

With the addition, \textit{post}-polymerization, of GA to our samples (see Methods), we change collagen's material properties (elastic modulus, plasticity threshold) without changing the microstructure of the network. For small deformations, the initial response is  similar to that of the unfixed sample, indicating that fiber alignment and the anomalous Poisson effect are only weakly sensitive to the fiber material properties. Furthermore, during cycles of applied deformation - i.e. ramping up the applied strain to 15\% and returning back to 0\% at the same rate - GA-crosslinked samples exhibit near-reversibility at the microscale, with fiber images of successive cycles being almost identical (figures \ref{fig:reversibility}H-J). Figures \ref{fig:reversibility}F and \ref{fig:reversibility}G quantify this reversibility in terms of the orientation histogram and the order parameter over multiple cycles. Consistent with this behavior, we find (see figure \ref{fig:reversibility}K) that the local deformation field at the mesoscale evolves along a reversible path for the same strain rate ($2.5 \cdot 10^{-4}$/s). Taken together, these results show that fiber plasticity, though observed for pure collagen samples, is unimportant in determining alignment at microscopic scales and the large Poisson ratio at larger scales. That is, strain-induced alignment is primarily an elastic effect.

The reversible behavior characterized above might then serve as a baseline to study more subtle effects (e.g. time-dependent and/or irreversible processes) that can be observed after many cycles or different strain rates. We see, in particular, a slight decrease in the amplitude of the alignment with the number of cycles, indicating that fiber plasticity still occurs, although on much larger timescales. In figure \ref{fig:reversibility}K we show the effects of strain rate on the response of the system for a fixed strain amplitude. At lower strain rates the system does not recover completely after a full cycle. This offset in the response can be attributed to a slow creeping process occurring over a time-scale of a few hours. At larger strain rates, one expects to see dynamic effects related to viscous dissipation. {Previous studies on the poroelasticity of collagen networks \cite{Wille2006, Knapp1997, Girton2002, Chandran2004,Ozerdem1995}, have reported equilibration times ranging from a few seconds to a few hours, probably due to the diversity of geometries and setups used for all these measurements. We find that dynamic effects associated with higher strain rates induce an asymmetry in the response between loading and unloading, influencing the unloading curve more than the loading curve. This shows that the material responds with different time-scales in extension and compression; this, in turn, suggests that different physical processes are involved during the loading which is dominated by fiber stretching and unloading which is dominated by fiber bending.}

\section*{Discussion}

In this study, we used an experimental and computational approach to quantify the emergence of fiber alignment as a collagen sample is stretched. This behavior is consistent with observations of cell-induced morphological changes in tissue equivalents and sheds light on a biologically relevant material non-linearity that arises from stress heterogeneities in fiber networks. Fiber alignment at the microscale results in tissue densification when boundary conditions allow it; this leads to high values of the Poisson ratio at large deformations, and is  observed for all studied collagen concentrations, with or without addition of glutaraldehyde.

The reversibility of fiber alignment and gel densification, seen in crosslinked collagen samples, show that these effects are primarily elastic.
Experiments on a piece of synthetic felt \cite{Kabla2007} have demonstrated that geometry alone can account for such a behavior based on the generic non-linearity of individual fibers, stiff in extension, but soft in compression (bending/buckling). Under uniaxial loading, a tensile stress is necessarily balanced in the microstructure by a compressive load on fibers normal to the stretch direction, leading to a collapse of the material in the normal direction and a strong enghancement of fiber alignment along the load direction, as observed here. This argument also explains the correlation between the moment where stress builds up (onset of non-linearity in rheological measurements) and the critical strain associated with the Poisson effect. This is also consistent with the bulk rheology experiments performed by {\it Janmey et al.} \cite{Janmey2007} who report large negative normal stresses in all tested fibrous materials. The normal stress in volume-constrained geometries (as in simple shear flows) is the counterpart of high Poisson ratios in unconstrained tests such as the local stretch performed in the present study.

Our findings seem to contrast with the experimental results of {\it Tower et al.} \cite{Tower2002} on stretched collagen samples, where either irreversible alignment (pure collagen sample) or early fracture with negligible alignment (GA crosslinked) occurs, suggesting that fiber plasticity is a key player in the alignment process. The difference with the present study can be readily explained from a geometrical standpoint:  {\it Tower et al.} used a geometry in which the width of the stretched region is comparable to its length; this might significantly constrain transverse motion of the material and prevent local volume changes to their full extent. By contrast, the 2-point stretching device we use ensures that the material is free to move along the direction normal to the load.

Our experiments have demonstrated the importance of sample geometry and boundary conditions on the microstructure and mechanical response of reconstituted biopolymer gels. For functional tissues, it is known that mechanical properties are often finely controlled by the texture of the underlying extracellular matrix as well \cite{Shadwick1999}. Understanding the mechanisms leading to such organization is  an important step in learning how it happens in the formation of natural tissues and for developing strategies to engineer suitable tissue equivalents. We have shown here that ECM texturization can be brought about simply by applying a deformation with a purely mechanical device, without any intervention due to active modeling by cells. This external perturbation is applied {\it post}-polymerization, in contrast with previous reports of fiber alignment induced by a flow \cite{Lee2006} or a magnetic field \cite{Barocas1998, Guo2007} {\it during} collagen polymerization. This provides direct support to the {\it in vivo} studies of {\it post}-polymerization collagen texturization in developing tissues \cite{Stopak1985}. However, the microscopic origin of the permanent texturization occurring in our uncrosslinked samples remains unclear. Fibril-fibril junctions are likely to be where plastic deformation occurs, allowing fibrils to slide with respect to one another and thus inducing irreversible changes in the network topology. One of the effects of glutaraldehyde crosslinking is to strengthen these junctions and reduce the amount of plastic reorganization allowed at the network level.

Whatever their origin, geometrical changes in the network structure are known to influence, in turn, cell behavior. Our mechanical setup allows for the dynamical control of network texture in a passive system, but clearly can be used to study how active cells respond to externally-induced anisotropy. Such experiments will provide more insights into specific mechanisms of mechanotransduction and cell behavior, which are crucial to processes such as morphogenesis, stem cell differentiation, metastasis and wound healing. More generally, the external control of fiber alignment \textit{post} polymerization not only offers a convenient way to design anisotropic tissue equivalents with collagen, but can also, when applied to other biopolymer systems, shed light on a range of analogous phenomena, such as actin gel contraction \cite{Bendix2008} or platelet-fibrin interactions \cite{Shah1997}, where microscopic agents interact through the network and lead to large scale evolution and reorganization of matter.

\footnotesize{
\bibliography{Fibers}
}
\end{document}